\documentclass[12pt]{iopart}

\usepackage{iopams,graphicx}
\begin{document}

\title[RSB in trajectory space for diffusion in logarithmically correlated random potentials]{Replica symmetry breaking in trajectory space for diffusion in logarithmically correlated random potentials}

\author{Masahiko Ueda}
\address{Department of Basic Science, The University of Tokyo, Tokyo 153-8902, Japan}
\ead{ueda@complex.c.u-tokyo.ac.jp}

\begin{abstract}
We study the dynamics of a particle in a one-dimensional Gaussian random potential with logarithmic correlations.
It was shown in previous studies that the model exhibits a dynamical transition between two subdiffusive phases.
We numerically investigate both phases by focusing on overlap between trajectories of two independent particles in a common random potential, and show that replica symmetry breaking in trajectory space occurs in the low-temperature phase.
\end{abstract}

\pacs{05.40.-a, 05.70.Fh, 64.70.P-}

\maketitle

\section{Introduction}
\label{sec:intro}
A particle in a Gaussian random potential with logarithmic correlations has frequently been studied in several contexts.
For example, in spatial dimension $d=2$, a single vortex in a XY spin model with random gauge disorder is described by this model \cite{NSKL1995, ChaFer1995, Tan1996, KorNat1996, Sch1997, CarLeD1998}.
Another example is two-dimensional Dirac fermions in a random magnetic field \cite{CMW1996, KMT1996, CCFGM1997}.
It is also regarded as an extended model of diffusion in random potentials such as the Sinai model (linearly correlated random potentials) \cite{Sin1982, BouGeo1990} to logarithmically correlated case.

The equilibrium state of the model has extensively been studied \cite{NSKL1995, ChaFer1995, Tan1996, KorNat1996, Sch1997, CMW1996, CCFGM1997, CarLeD2001, FyoBou2008a, FyoBou2008b, FLR2009, FHK2012}.
The most intriguing result is that the model exhibits a localization phase transition.
In particular, the equilibrium (Boltzmann-Gibbs) distribution in the low-temperature phase is dominated by a few distant states and described by replica symmetry breaking (RSB) \cite{CarLeD2001}, similarly to mean-field spin glass models \cite{MPV1987}.
This is in contrast to the result for the Sinai model, where a single state is dominant.
Furthermore, properties of the localized state is similar to those of directed polymers on the Cayley tree \cite{DerSpo1988}, rather than the random energy model (REM) \cite{Der1981}, as discussed in \cite{Tan1996, CMW1996, CarLeD2001}.
The extreme-value statistics is described by a non-Gumbel distribution \cite{CarLeD2001, FyoBou2008b}, differently from the Gumbel distribution for REM \cite{BouMez1997}.

The dynamics of the model on a finite-dimensional lattice has also been studied.
According to a renormalization group analysis, subdiffusion occurs in finite temperature \cite{BCGL1987, HonKar1988, DHP1990}.
Furthermore, the existence of a dynamical transition between two subdiffusive phases was proved in $d=1,2$ \cite{CasLeD2001}.
Particularly, the dynamical transition temperature coincides with the static transition temperature in $d=1$.
Although calculation of the mean first passage time in one dimension was exactly mapped into an equilibrium statistical mechanical problem in the proof of the existence of the dynamical transition, general relations between statics and dynamics are not known.
Especially, dynamical properties of the low-temperature phase beyond the single-particle diffusion law are not clear.

In this paper, we study the dynamics of independent particles in a common random potential in $d=1$.
Such situation has been investigated in the context of relative diffusion \cite{KlyTat1974, Deu1985, WilMeh2003, Ued2016, Gol1984, HalZha1995, Chi2002}.
In many cases, relative diffusion is qualitatively different from single-particle diffusion.
In our previous papers, we have developed the method to detect localization in trajectory space by using overlap between trajectories \cite{UedSas2015, UedSas2017}.
When relative diffusion is strongly suppressed, the overlap takes a nontrivial value.
Here, we apply this method to diffusion in logarithmically correlated random potentials, and numerically show that replica symmetry breaking in trajectory space occurs in the relaxation process in the low-temperature phase, which implies that a diffusion trajectory is localized into a few specific trajectories.

The paper is organized as follows.
In section \ref{sec:model}, we introduce the model and overlap between trajectories.
In section \ref{sec:results}, we provide numerical results supporting that RSB in trajectory space occurs in the low-temperature phase, while there is no RSB in the high-temperature phase.
Section \ref{sec:conclusion} is devoted to concluding remarks.

\section{Model}
\label{sec:model}
We consider a particle on a one-dimensional lattice of size $L$.
The position of the particle at time $t$ is denoted by $x(t) \in \left\{ 1, \cdots, L \right\}$.
The periodic boundary condition is imposed to the lattice.
For each site, a quenched random potential $V(x)$ is defined.
The dynamics of the particle is described by a continuous-time Markov jump process with the transition rate matrix
\begin{eqnarray}
 W(x\rightarrow x^\prime) &\equiv& e^{-\frac{1}{2}\beta \left\{ V(x+1) - V(x) \right\}} \delta_{x^\prime, x+1} + e^{-\frac{1}{2}\beta \left\{ V(x-1) - V(x) \right\}} \delta_{x^\prime, x-1},
 \label{eq:transition_rate}
\end{eqnarray}
where $\delta_{i,j}$ is the Kronecker delta.
Here, the parameter $\beta$ represents the inverse temperature, and we define temperature by $T\equiv 1/\beta$.
Below we denote the thermal average (average with respect to the probability distribution of particle trajectories) and the disorder average (average with respect to the probability distribution of $V$) by $\left\langle \cdots \right\rangle$ and $\mathbb{E}\left[ \cdots \right]$, respectively.
When we define the probability distribution of the particle position as $P(x,t)=\left\langle \delta_{x,x(t)} \right\rangle$, its time evolution is described by the master equation
\begin{eqnarray}
 \frac{\partial }{\partial t}P(x,t) &=& W(x-1\rightarrow x)P(x-1,t) + W(x+1\rightarrow x)P(x+1,t) \nonumber \\
 && \qquad - \left[ W(x\rightarrow x-1) + W(x\rightarrow x+1) \right]P(x,t).
\end{eqnarray}
The initial condition is denoted by $P_0(x)$.
Because of the detailed balance condition, the master equation has a stationary solution
\begin{eqnarray}
 P_\mathrm{eq}(x) &\equiv& \frac{1}{Z_\mathrm{eq}} e^{-\beta V(x)}
\end{eqnarray}
with $Z_\mathrm{eq} \equiv \sum_{x=1}^L e^{-\beta V(x)}$ for finite $L$.
We consider the case that the random potential $V(x)$ is a logarithmically correlated Gaussian random variable with zero mean:
\begin{eqnarray}
 \mathbb{E}\left[ V(x) \right] &=& 0 \\
 \mathbb{E}\left[ \left\{ V(x)-V(x^\prime) \right\}^2 \right] &\sim& 4\log\left| x-x^\prime \right|.
\end{eqnarray}

We briefly review the dynamical properties of the model with $L\rightarrow \infty$.
In the previous studies \cite{BCGL1987, HonKar1988, DHP1990}, it was shown that subdiffusion is observed in the system.
Furthermore, in the previous study \cite{CasLeD2001}, it was proved that the system exhibits a dynamical transition between two subdiffusive phases:
\begin{eqnarray}
 \mathbb{E}\left[ \left\langle \left\{ x(t)-x(0) \right\}^2 \right\rangle \right] &\sim& \left\{
\begin{array}{ll}
 t^\frac{T^2}{T^2+1}  &\quad (T>1) \\
 t^\frac{T}{2} &\quad (T<1).
\end{array}
\right.
\label{eq:subdif}
\end{eqnarray}
Although a non-equilibrium splitting of the thermal distribution of the diffusing particle into a few packets in the low temperature phase was also suggested in \cite{CasLeD2001}, it has not been detected explicitly.

Here, we investigate both the high-temperature phase and the low-temperature phase by mainly focusing on the behavior of particles on a common random potential with $L\rightarrow \infty$.
In particular, we study the probability distribution of overlap between two trajectories.
We consider two independent particles on a common random potential.
Overlap of the two particle trajectories $x^{(1)}(t)$ and $x^{(2)}(t)$ is defined as
\begin{eqnarray}
 q(t) &=& \frac{1}{t} \int_0^t dt^\prime \delta_{x^{(1)}(t^\prime), x^{(2)}(t^\prime)}.
\end{eqnarray}
This quantity describes the similarity of two trajectories to each other.
When trajectories of particles are localized in trajectory space, $q$ takes a finite value even in $t\rightarrow \infty$.
In contrast, when trajectories are not localized, $q$ takes zero.
The average overlap $\mathbb{E}\left[ \left\langle q \right\rangle \right]$ is useful to detect the existence of localization in trajectory space.
It should be noted that the average overlap is related to the participation ratio $Y_2(t)\equiv \sum_x P(x,t)^2$ \cite{Der1997} by
\begin{eqnarray}
 \left\langle q(t) \right\rangle &=& \frac{1}{t} \int_0^t dt^\prime Y_2(t^\prime),
 \label{eq:q-Y}
\end{eqnarray}
and therefore $\left\langle q(t) \right\rangle$ and $Y_2(t)$ contain equal information.
Here, we focus on the overlap distribution
\begin{eqnarray}
 P(q) &=& \mathbb{E}\left[ \left\langle \delta\left( q - \frac{1}{t} \int_0^t dt^\prime \delta_{x^{(1)}(t^\prime), x^{(2)}(t^\prime)} \right) \right\rangle \right],
 \label{eq:ovdist}
\end{eqnarray}
which contains more information than the participation ratio.
The distribution of overlap was originally introduced in spin glass theory in order to detect the existence of several stable spin configurations in a spin glass phase \cite{MPV1987}.
When $P(q)$ has only one trivial peak $\delta(q)$ in $t\rightarrow \infty$, there is no localization.
When $P(q)$ has one nontrivial peak $\delta(q-q_*)$ with $q_*\neq 0$, the system is localized into one stable trajectory.
When $P(q)$ is a nontrivial function with a trivial peak $\delta(q)$, the system has several stable trajectories.
In the last case, replica symmetry in trajectory space is said to be broken.

\section{Numerical results}
\label{sec:results}
We generate random variables $\left\{ V(x) \right\}$ according to the method proposed in Ref. \cite{CarLeD2001}.
The potential $V(x)$ is computed from its Fourier components
\begin{eqnarray}
 V(x) &=& w_\frac{L}{2}(-1)^x + \sum_{k=1}^{\frac{L}{2}-1}w_k \cos\left( \frac{2\pi k x}{L} -\phi_k \right),
\end{eqnarray}
where $w_k$ is a zero-mean independent Gaussian random variable with $\mathbb{E}\left[ w_kw_{k^\prime} \right] = \Delta(k)\delta_{k,k^\prime}$, and $\phi_k$ independently obeys the uniform distribution in $[0,2\pi]$.
Here, $\Delta(k)$ is chosen as
\begin{eqnarray}
 \Delta(k) &=& \left\{
\begin{array}{ll}
 \frac{4\pi}{L} \frac{1}{\left| \sin\left( \frac{\pi k}{L} \right) \right| \sqrt{ 6 - 2\cos\left( \frac{2\pi k}{L} \right) }}  &\quad \left( k=1,\cdots, \frac{L}{2}-1 \right) \\
 \frac{2\pi}{L} \frac{1}{2\sqrt{2}} &\quad \left( k=\frac{L}{2} \right).
\end{array}
\right.
\end{eqnarray}
By this choice, $\mathbb{E}\left[ \left\{ V(x)-V(x^\prime) \right\}^2 \right] \sim 4\log\left| x-x^\prime \right|$ is realized in $1\ll \left| x-x^\prime \right| \ll L/2$.

We perform numerical simulations for $N$ particles on a common random potential.
We set the system size $L$ as $L=10000$.
When time $t$ satisfies $t\ll \tau_\mathrm{eq}(L)$, where $\tau_\mathrm{eq}(L)$ is an equilibration time, the system size $L$ is effectively regarded as infinity.
Here $\tau_\mathrm{eq}(L)$ is estimated from the time needed for the particle to diffuse a distance $L$ by subdiffusion (\ref{eq:subdif}).
We discretize time $t$ with the width $\Delta t=0.01$.
The initial condition is set to $P_0(x)=\delta_{x,\frac{L}{2}}$.
The number of particles for calculating the thermal average is set to $N=10000$, and the disorder average is calculated from $10000$ samples.

\begin{figure}[t]
\includegraphics[clip, width=8.0cm]{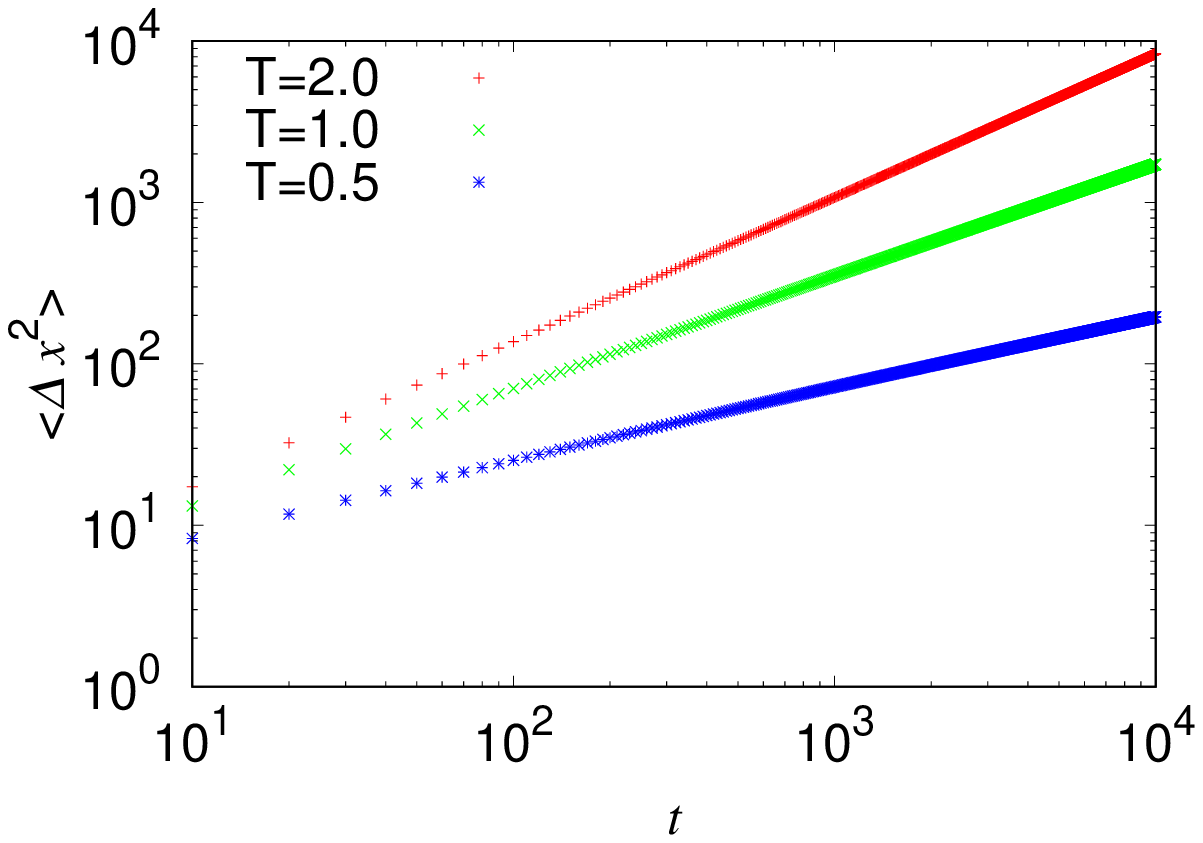}
\includegraphics[clip, width=8.0cm]{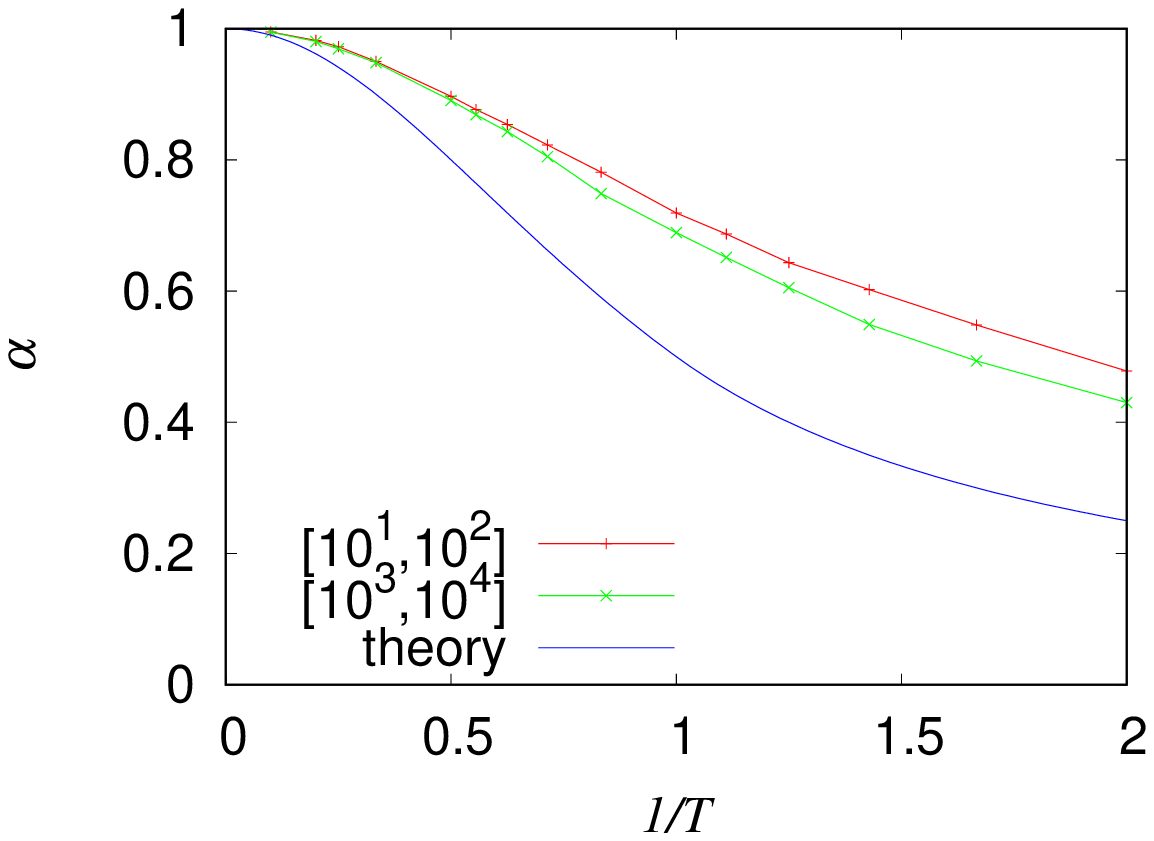}
\caption{(Left) The mean-squared displacement $\mathbb{E}\left[ \left\langle \left\{ x(t)-x(0) \right\}^2 \right\rangle \right]$ for various $T$ as a function of time $t$ in a log-log plot. (Right) Temperature $T$ dependence of the exponent $\alpha$ with $\mathbb{E}\left[ \left\langle \left\{ x(t)-x(0) \right\}^2 \right\rangle \right] \sim t^\alpha$. The guideline corresponds to the theoretical prediction (\ref{eq:subdif}).}
\label{fig:xv}
\end{figure}

First, we numerically calculate the single-particle mean-squared displacement $\mathbb{E}\left[ \left\langle \left\{ x(t)-x(0) \right\}^2 \right\rangle \right]$ in order to check the previous result (\ref{eq:subdif}).
The mean-squared displacement at temperature $T=2.0$, $1.0$, $0.5$ is displayed in the left side of Fig. \ref{fig:xv}.
Subdiffusion $\mathbb{E}\left[ \left\langle \left\{ x(t)-x(0) \right\}^2 \right\rangle \right] \sim t^\alpha$ $(\alpha<1)$ is observed for all cases.
Moreover, we estimate the exponent $\alpha$ for various $T$ by fitting, which is plotted in the right side of Fig. \ref{fig:xv}.
The fitting is done by using numerical data in the ranges $t\in \left[ 10^1, 10^2 \right]$ and $t\in \left[ 10^3, 10^4 \right]$.
The guideline corresponds to the theoretical value (\ref{eq:subdif}).
Although numerical data do not quantitatively agree with the theoretical prediction, the numerical data becomes closer to (\ref{eq:subdif}) as $t$ increases.
The deviation from the theoretical value comes from the fact that $t$ is not large enough.

\begin{figure}[t]
\includegraphics[clip, width=8.0cm]{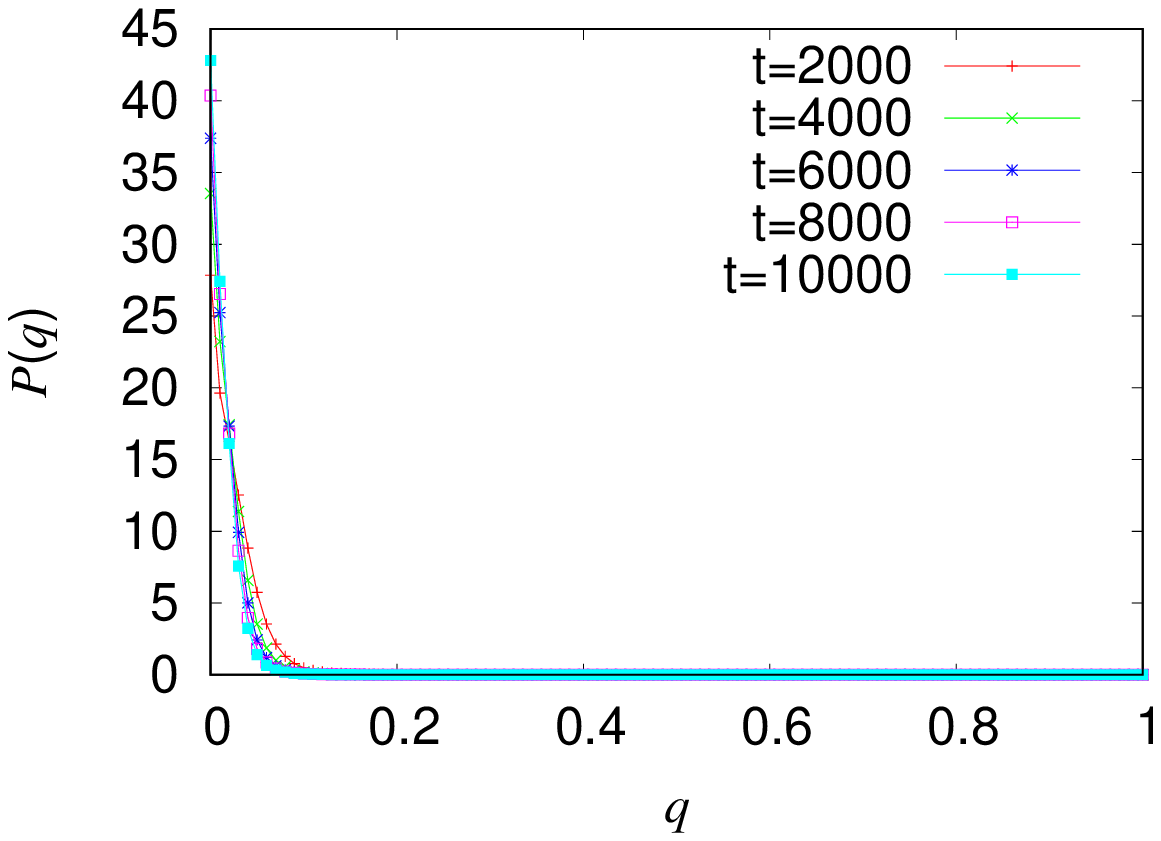}
\includegraphics[clip, width=8.0cm]{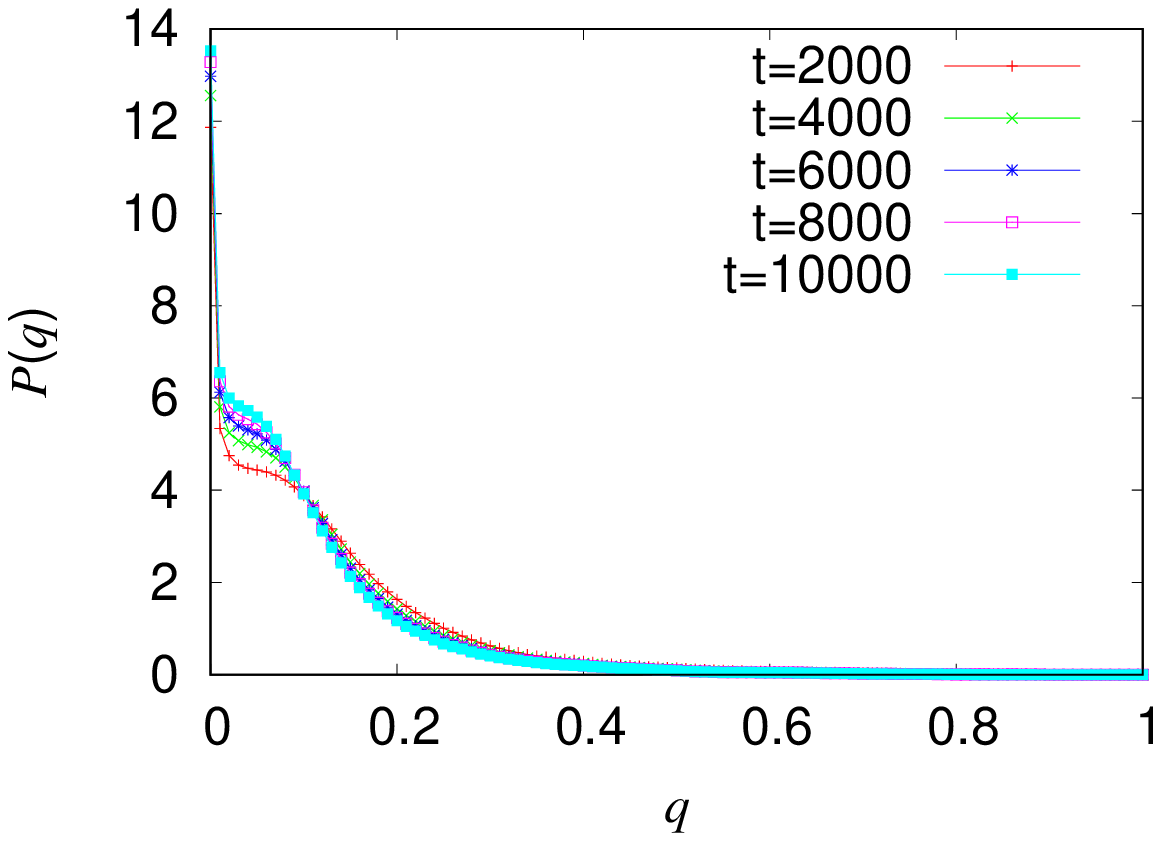}
\includegraphics[clip, width=8.0cm]{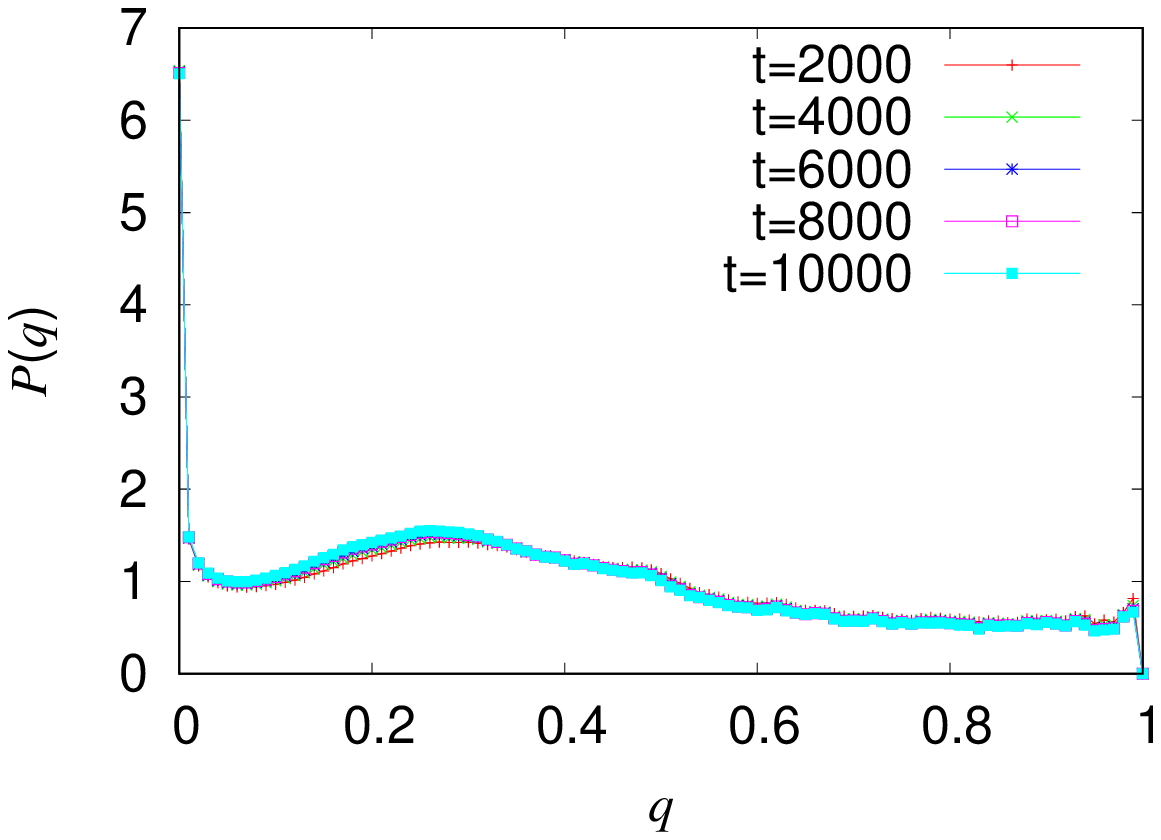}
\caption{The distribution of ovelap $P(q)$ for $T=2.0$ (left top), $T=1.0$ (right top) and $T=0.5$ (left bottom).}
\label{fig:ovhist}
\end{figure}

Next, we calculate the distribution of overlap (\ref{eq:ovdist}).
$P(q)$ for $T=2.0$, $1.0$, $0.5$ is displayed in Fig. \ref{fig:ovhist}.
When $T>1$, $P(q)$ relaxes to the trivial function $\delta(q)$, where the peak $P(q=0)$ increases with time.
In contrast, when $T<1$, $P(q)$ seems to converge to some nontrivial function with finite $P(q=0)$.
At the dynamical transition temperature $T=1$, there exists a precursor of a nontrivial peak, which will be absorbed into the peak at $q=0$ with time.
Therefore, we conclude that RSB in trajectory space occurs in the low-temperature phase $T<1$, while there is no RSB in the high-temperature phase $T>1$.
We remark that the form of $P(q)$ at $T<1$ is similar to that observed in the relaxation process of the trap model \cite{UedSas2017}.
It seems to be different from a standard one-step RSB-type distribution with two delta peaks.
This result leads to the expectation that dynamical overlap studied here obeys a different statistical law from that of static overlap, which obeys a 1RSB-type distribution \cite{CarLeD2001}.

\begin{figure}[t]
\includegraphics[clip, width=8.0cm]{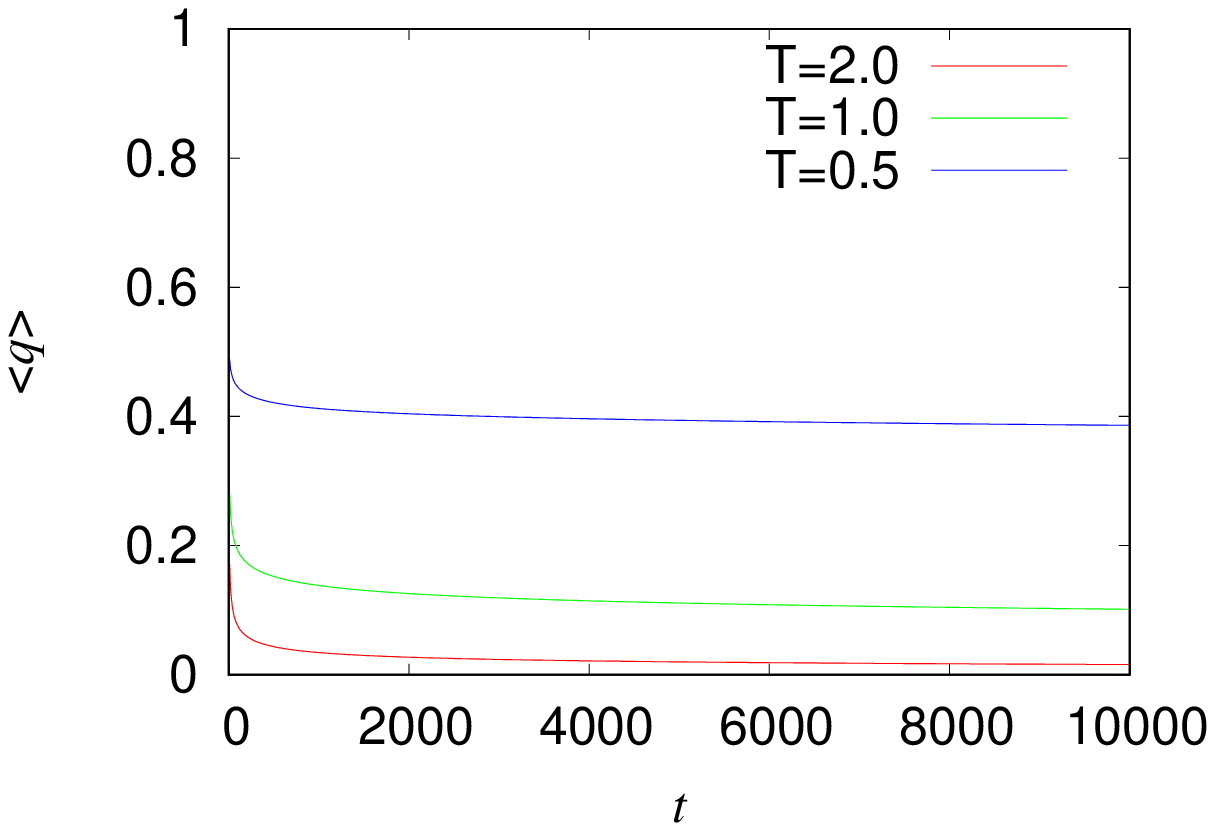}
\includegraphics[clip, width=8.0cm]{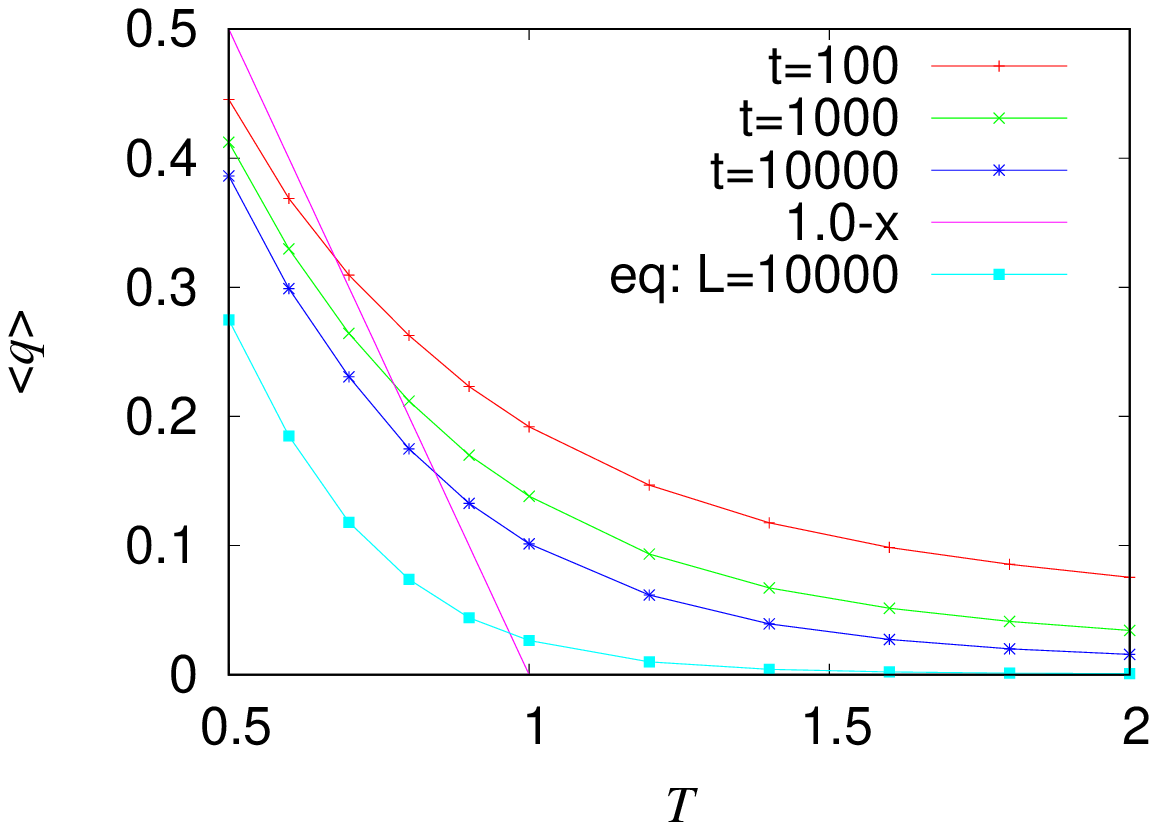}
\caption{(Left) Time $t$ dependence of the expectation of overlap $\mathbb{E}\left[ \left\langle q(t) \right\rangle \right]$ for various $T$. (Right) Temperature $T$ dependence of the expectation of overlap $\mathbb{E}\left[ \left\langle q(t) \right\rangle \right]$ for various $t$. The solid line corresponds to (\ref{eq:Yeq_cal}). A numerical value of $\mathbb{E}\left[ Y_2^{(\mathrm{eq})} \right]$ for $L=10000$ is also displayed.}
\label{fig:ov}
\end{figure}

We further calculate the expectation of overlap $\left\langle q \right\rangle$.
In the left side of Fig. \ref{fig:ov}, the time evolution of the average overlap $\mathbb{E}\left[ \left\langle q(t) \right\rangle \right]$ is displayed.
We can see the decay of $\mathbb{E}\left[ \left\langle q(t) \right\rangle \right]$ in $T>1$, which is consistent with the result from $P(q)$.
In $T<1$, $\mathbb{E}\left[ \left\langle q(t) \right\rangle \right]$ seems to converge to a finite value, although relaxation is very slow.
We also display the temperature dependence of $\mathbb{E}\left[ \left\langle q(t) \right\rangle \right]$ for various $t$ in the right side of Fig. \ref{fig:ov}.
Because of the relation (\ref{eq:q-Y}), the large $t$ limit of $\mathbb{E}\left[ \left\langle q(t) \right\rangle \right]$ is equivalent to the large $t$ limit of $\mathbb{E}\left[ Y_2(t) \right]$.
In general, the order of the two limits $L\rightarrow \infty$ and $t\rightarrow \infty$ is crucial in the calculation of $\mathbb{E}\left[ Y_2(t) \right]$.
Note that
\begin{eqnarray}
 \lim_{L\rightarrow \infty} \lim_{t\rightarrow \infty} \mathbb{E}\left[ Y_2(t) \right] &=& \lim_{L\rightarrow \infty} \mathbb{E}\left[ Y_2^{(\mathrm{eq})} \right]
\end{eqnarray}
with $Y_2^{(\mathrm{eq})} \equiv \sum_x P_\mathrm{eq}(x)^2$.
In \cite{CarLeD2001}, it was discussed that
\begin{eqnarray}
 \lim_{L\rightarrow \infty} \mathbb{E}\left[ Y_2^{(\mathrm{eq})} \right] &=& \left\{
\begin{array}{ll}
 0 &\quad (T>1) \\
 1-T &\quad (T<1)
\end{array}
\right.
 \label{eq:Yeq_cal}
\end{eqnarray}
as in the standard 1RSB case.
In our numerical simulation, another limit $\lim_{t\rightarrow \infty} \lim_{L\rightarrow \infty} \mathbb{E}\left[ Y_2(t) \right]$ is computed.
In the right side of Fig. \ref{fig:ov}, the theoretical result (\ref{eq:Yeq_cal}) together with a numerical value of $\mathbb{E}\left[ Y_2^{(\mathrm{eq})} \right]$ for $L=10000$ is also displayed.
We can see finite size effect for the equilibrium participation ratio $\mathbb{E}\left[ Y_2^{(\mathrm{eq})} \right]$.
At this stage, we cannot conclude whether the two limits $\lim_{L\rightarrow \infty} \lim_{t\rightarrow \infty} \mathbb{E}\left[ Y_2(t) \right]$ and $\lim_{t\rightarrow \infty} \lim_{L\rightarrow \infty} \mathbb{E}\left[ Y_2(t) \right]$ are different, as in the one-dimensional trap model \cite{BerBou2003}.
This problem is beyond the scope of this paper, and further investigation will be made in future.

\section{Conclusion}
\label{sec:conclusion}
In this paper, we studied one-dimensional diffusion of particles in a common Gaussian random potential with logarithmic correlations.
We numerically calculated the probability distribution of overlap between trajectories of two independent particles, and found that replica symmetry breaking in trajectory space occurs in the low-temperature phase, implying that a diffusion trajectory freezes into several stable trajectories, while there is no RSB in the high-temperature phase.
The type of RSB seems not to be 1RSB, which is different from results for the localization of states in statics.
Developing some theoretical techniques to detect RSB in trajectory space is a future work.

We remark on the difference between our results and ``non-equilibrium 1-step RSB'' conjectured in \cite{CasLeD2001}.
In \cite{CasLeD2001}, it was suggested that a non-equilibrium splitting of the thermal distribution of the diffusing particle into a few packets occurs in the low-temperature phase and this phenomenon is described by a non-equilibrium 1-step RSB.
The meaning of their ``non-equilibrium 1-step RSB'' is as follows.
First, they define the mean first passage time $t_1$ from $x=0$ to $x=L$, which is similar to the partition function of two copies.
The dynamical exponent is then obtained by
\begin{eqnarray}
 z &=& \lim_{L\rightarrow \infty} \frac{1}{\log L} \mathbb{E}\left[ \log t_1 \right].
\end{eqnarray}
This quantity is similar to the free energy of two copies.
The disorder average $\mathbb{E}\left[ \log t_1 \right]$ is calculated by the replica method.
They found that this quantity in the low-temperature phase is calculated by the 1RSB saddle point.
This implies that the mean first passage time is dominated by the crossing time for a few valleys and hills.
This result is about a state of the system.
In contrast, we study overlap between trajectories.
A trajectory is the time series of a state, and includes more information than a state itself.
Therefore, the statistics of overlap between trajectories does not necessarily coincide with overlap between states.

Before ending this paper, we compare the result of this paper with the previous results on RSB in trajectory space \cite{UedSas2015, UedSas2017}.
RSB in trajectory space has been observed in two models, a tracer particle on a one-dimensional KPZ field and the one-dimensional quenched trap model.
In the former case, the stationary probability distribution of a KPZ field is equal to the probability distribution of a potential in the Sinai model, which describes diffusion in a time-independent random potential with linear correlations $\mathbb{E}\left[ \left\{ V(x)-V(x^\prime) \right\}^2 \right] = \left| x-x^\prime \right|$.
As mentioned in section \ref{sec:intro}, particles on a Sinai landscape are collected into a single valley \cite{Gol1984}, and therefore the Sinai model is not expected to exhibit RSB in trajectory space.
However, because a KPZ field is time-dependent, properties of the model are different from those of the Sinai model, and several dominant valleys appear in trajectory space, which leads to RSB in trajectory space.
In contrast, in the present model, RSB in trajectory space is realized by weakening correlations of potentials from linear one to logarithmic one.
In the case of the trap model, diffusion in a time-independent potential is considered, similarly to the model in this paper.
In statics, the trap model is a kind of REM \cite{BouMez1997}.
In contrast, as mentioned in section \ref{sec:intro}, the present model is qualitatively different from REM even in statics because potential energy is correlated.
Therefore, we believe that the result of this paper is nontrivial compared to those of the previous studies.

\ack
The present study was supported by JSPS KAKENHI Grant Numbers JP16J00178.

\end{document}